\documentclass{iopart}

\usepackage{iopams}

\expandafter\let\csname equation*\endcsname\relax
\expandafter\let\csname endequation*\endcsname\relax

\usepackage{amsmath}

\usepackage{graphicx}% Include figure files
\usepackage[utf8]{inputenc}
\usepackage{color}

\usepackage{cite}

\newcommand{\mev}{\,{\rm MeV}}

\begin{document}

{\vbox{\hbox{ADP-19-6/T1086}}
{\vbox{\hbox{LTH 1200}}
{\vbox{\hbox{DESY 19-053}}

\title[Isospin splittings in decuplet baryons]{Isospin splittings in the decuplet baryon spectrum from dynamical
  QCD+QED}

\author{R.~Horsley$^1$,
  Z.~Koumi$^2$,
  Y.~Nakamura$^3$,
  H.~Perlt$^4$,
  D.~Pleiter$^{5,6}$,
  P.E.L.~Rakow$^7$,
  G.~Schierholz$^8$,
  A.~Schiller$^4$,
  H.~St\"uben$^9$,
  R.D.~Young$^2$ and J.M.~Zanotti$^2$}

\address{$^1$ School of Physics and Astronomy, University of Edinburgh,
  Edinburgh EH9 3FD, UK}
\address{$^2$ CSSM, Department of Physics, University of Adelaide,
   SA, Australia}
\address{$^3$ RIKEN Center for Computational Science,
   Kobe, Hyogo 650-0047, Japan}
\address{$^4$ Institut f\"ur Theoretische Physik, Universit\"at
   Leipzig, 04109 Leipzig, Germany}
\address{$^5$ J\"ulich Supercomputer Centre, Forschungszentrum
   J\"ulich, 52425 J\"ulich, Germany}
\address{$^6$ Institut f\"ur Theoretische Physik, Universit\"at
   Regensburg, 93040 Regensburg, Germany}
\address{$^7$ Theoretical Physics Division, Department of Mathematical
   Sciences, University of Liverpool, Liverpool L69 3BX, UK}
\address{$^8$ Deutsches Elektronen-Synchrotron DESY, 22603 Hamburg,
   Germany}
\address{$^9$ Regionales Rechenzentrum, Universit\"at Hamburg, 20146
  Hamburg, Germany}
\address{}
\address{CSSM/QCDSF/UKQCD Collaboration}

\date{\today}

 \begin{abstract}
We report a new analysis of the isospin splittings within the decuplet
baryon spectrum.
Our numerical results are based upon five ensembles of dynamical
QCD+QED lattices.
The analysis is carried out within a flavour-breaking
expansion which encodes the effects of breaking the quark masses and
electromagnetic charges away from an approximate SU(3) symmetric point.
The results display total isospin splittings within the approximate
SU(2) multiplets that are compatible with phenomenological estimates.
Further, new insight is gained into these splittings by separating the
contributions arising from strong and electromagnetic effects.
We also present an update of earlier results on the octet baryon spectrum.
\end{abstract}

%\pacs{Valid PACS appear here}% PACS, the Physics and Astronomy
%                             % Classification Scheme.
%\keywords{Suggested keywords}%Use showkeys class option if keyword
%                              %display desired
\maketitle
%
%%%%%
%%%%%%%%%%%%%%%%%%%%%%%%%%%%%%%%%%%%%%%%%%%%%%%%%%%%%%%
%%%%%
%
\section{Introduction}
\label{sec:intro}
Isospin symmetry is so prevalent in the description of hadronic
systems that the significance of a potential violation of this
symmetry is often taken for granted.
In the context of spectroscopy, isospin symmetry is manifest in the
approximate degeneracy of isospin multiplets.
Using this approximate symmetry to inform the interpretation of states
can be incredibly powerful.
Nevertheless, violations of this symmetry will become significant at some degree of precision.
Isospin splittings in the ground state hyperons are known to be as
much as $8\,{\rm MeV}$.
Phenomenological estimates suggest splittings in the decuplet baryons
to be of a similar size \cite{Cutkosky:1992nx}.
Given that modern analyses are able to achieve (real part of the) pole
positions at a precision of $\pm 1\,{\rm MeV}$ precision for the
$\Delta$-baryons \cite{Svarc:2014zja,Patrignani:2016xqp}, there is an
opportunity to revisit analyses of isospin violation in low-energy
$\pi N$ scattering \cite{Gibbs:1995dm,Matsinos:1997pb}.

Beyond spectroscopy, it is also worth noting that the determination of
isospin violation is relevant to a range of physical phenomena,
including the flavour decomposition of nucleon structure
\cite{Miller:2006tv,Londergan:2009kj,Wagman:2014nfa,Shanahan:2015caa};
tests of neutrino-nucleus interactions
\cite{Zeller:2001hh,Bentz:2009yy}; precision constraints on CKM
\cite{Cabibbo:1963yz,Kobayashi:1973fv} matrix elements from leptonic
\cite{Cirigliano:2011tm,Lucha:2016nzv} and semi-leptonic
\cite{Cirigliano:2004pv} decay rates; and quark mass parameters
\cite{Gasser:1982ap,Leutwyler:1996sa,Gasser:2003hk,Colangelo:2016jmc}.
In addition, the interplay of the coupled gauge theories in the
nonperturbative domain offers a unique theoretical playground to
explore.
These extended motivations have prompted intensive effort in recent
years to introduce electromagnetic effects in numerical lattice QCD
studies
\cite{Blum:2010ym,Aoki:2012st,deDivitiis:2013xla,Borsanyi:2013lga,Borsanyi:2014jba,Endres:2015gda,Horsley:2015eaa,Horsley:2015vla,Giusti:2017dmp,Boyle:2017gzv,Giusti:2017dwk}
--- building upon the pioneering work of Duncan, Eichten \& Thacker
\cite{Duncan:1996xy}.

In the present work, we perform simulations in dynamically-coupled
QCD+QED \cite{Horsley:2015eaa,Horsley:2015vla}, where the electric
charges of sea-quark loops are included in the fermion determinant.
In this work, the hadron spectrum calculations are performed across $32^3\times 64$
and $48^3\times 96$ lattices with up to 3 distinct sea quark mass
combinations.
Partially-quenched correlators are employed to further 
constrain flavour symmetry breaking effects.
Starting from an SU(3) symmetric point inspired by Dashen's relation
\cite{Dashen:1969eg}, we use a flavour symmetry breaking expansion
\cite{Horsley:2015vla} to extrapolate to the physical quark masses and
interpolate to the physical QED coupling --- where our underlying
gauge ensembles use an unphysically-large $\alpha_{\rm QED}\sim 0.1$
to enhance the signal strength in the electromagnetic effects.
In addition to providing isospin splittings among the decuplet
multiplets, we also present updated results for the octet baryons.

The manuscript proceeds as follows: Section II reviews the form of the
flavour-symmetry breaking expansions, including a description of the
``Dashen scheme'' used to distinguish electromagnetic and quark mass
effects; Section~\ref{sec:lattice matters} provides the lattice simulation
details; Section~\ref{sec:determining} presents the lattice spectra results,
including the flavour-breaking fits and treatment of finite-size
effects; results and discussion follow in Section~\ref{sec:results}; and we
conclude in Section~\ref{sec:conclusion}.

%
%%%%%
%%%%%%%%%%%%%%%%%%%%%%%%%%%%%%%%%%%%%%%%%%%%%%%%%%%%%%%
%%%%%
%

\section{Mass expansions}
\label{sec:massexpansion}
The approximate SU(3) flavour symmetry of nature has provided
tremendous insight into strong interaction phenomenology.
In recent lattice studies of pure QCD, we have exploited this symmetry
by formulating an SU(3) expansion about a point of exact flavour
symmetry \cite{Bietenholz:2011qq}.
The key to these investigations has been to use a starting point where
the degenerate light (up, down and strange) quark mass is
approximately equal to the average of the corresponding physical
masses, $\bar{m} = (m_u + m_d + m_s)/3$.
As a consequence, when quark masses are tuned to lie on a trajectory
that holds $\bar{m}$ fixed at its physical value, flavour-singlet
quantities only vary at second order in the dominant SU(3) breaking
parameter $\delta m_q = m_q-\bar{m}$.
This particular value is chosen such that lattice determinations of
flavour-singlet quantities, such as $X_\pi^2 = (2m_K^2 + m_\pi^2)/3$,
take their physical value.
The extrapolation to the physical point along a trajectory with
$\bar{m}=\mathtt{constant}$ is simplified by the reduced set of
operators that contribute to the quark mass variation
\cite{Bietenholz:2011qq}.
Isospin violating effects arising from the quark mass difference
$m_d-m_u$ are also naturally incorporated into the formulation
\cite{Horsley:2012fw}.

Upon inclusion of electromagnetic effects, we wish to further exploit
the perturbative breaking of SU(3) symmetry.
The electromagnetic renormalisation of the quark masses makes it
impossible to rigorously define equality of the light quark masses
$m_u=m_d$ in a scheme-invariant fashion.
Nevertheless, by choosing an appropriate renormalisation condition, we
can ensure that electromagnetic effects can be treated perturbatively.
Inspired by the Dashen relation \cite{Dashen:1969eg}, we impose the
condition that the QCD component of neutral pseudoscalar mesons at the
symmetric point can be parameterised identically and hence are
equal.
In practice, our tuning procedure requires that the bare quark masses,
$m_q$ (or $\kappa_q$ in the case of Wilson fermions), at the SU(3)
symmetric point are chosen such that all neutral (connected) pseudoscalar mesons
$M(q\bar{q})$ are equal, i.e. $M^2(u\bar{u})\approx
M^2(d\bar{d}) = M^2(s\bar{s})$, where equality between
$M^2(d\bar{d})$ and $M^2(s\bar{s})$ at the symmetric point is
exact due to the fact that $d$ and $s$ quarks have the same charge.
Further details of the Dashen scheme and the associated tuning can be
found in Ref.~\cite{Horsley:2015vla}.

Following the procedure outlined in Ref.~\cite{Bietenholz:2011qq},
adapted to incorporate electromagnetic corrections
\cite{Horsley:2015vla}, we obtain the relevant flavour-breaking
expressions for our hadron masses.
The flavour-breaking expansion of the pseudoscalar masses to NLO was
reported in Ref.~\cite{Horsley:2015vla}, however we quote the result
here for completeness, albeit with a slight rearrangement of the terms
 \begin{align}
M^2(a \bar b) =& M_0^2 + \alpha ( \delta \mu_a + \delta \mu_b) + \beta_1 ( \delta \mu_a^2 + \delta \mu_b^2)   \nonumber
\\&+ \beta_2 ( \delta \mu_a - \delta \mu_b)^2  + \beta_1^{EM} (e_a^2 + e_b^2) + \beta_2^{EM} (e_a- e_b)^2  \nonumber
\\&+\gamma_1^{EM} ( e_a^2 \delta \mu_a + e_b^2 \delta \mu_b ) +\gamma_2^{EM} (e_a e_b) (\delta \mu_a + \delta \mu_b ) \nonumber 
\\&+\gamma_3^{EM} ( e_b^2\delta \mu_a + e_a^2\delta \mu_b ) \nonumber
\\&+c_1 (\delta m_u + \delta m_d + \delta m_s) \nonumber
\\&+c_2 \left[ \delta m_u^2 + \delta m_d^2 + \delta m_s^2 -(\delta m_u \delta m_d +\delta m_u \delta m_s+\delta m_d \delta m_s)\right]\nonumber
\\&+c_3 (\delta m_u+ \delta m_d +\delta m_s)^2 +c_4 ( e_u^2 \delta m_u + e_d^2 \delta m_d + e_s^2 \delta m_s ) \nonumber
\\&+ c_1^{EM} (e_u^2 + e_d^2 + e_s^2) + c_2^{EM} (e_ue_d + e_ue_s + e_de_s) \nonumber
\\&+c_3^{EM} (e_u^2 +e_d^2 + e_s^2) (\delta \mu_a + \delta \mu_b).
\label{eq:meson_expansion_full} 
\end{align}
In this expansion, the valence quark charges are indicated by
$e_{a,b}$ and the sea quark charges by $e_{u,d,s}$.
The valence and sea quark mass deviations from the SU(3) symmetric
point are respectively denoted by
\begin{align}
  \delta\mu_{a,b}&=\mu_{a,b}-\bar{m},\quad
  \delta m_{u,d,s}=m_{u,d,s}-\bar{m}.
\end{align}
These quark mass variations are evaluated in the Dashen scheme
\cite{Horsley:2015vla}, where the distance from the symmetric point to
the chiral limit, $m^{sym}_q$, is defined to be independent of the
quark charge, hence absorbing the quark electromagnetic self energy
into the quark mass parameter.

Given that our framework is to approach the physical point along a
trajectory that holds the singlet quark mass approximately
constant\footnote{Note that in pure QCD the singlet quark mass can be
  held constant exactly, but once electromagnetism is included, this
  is only approximately true due to the different quark charges.}, we
can neglect the $c_1$ and $c_3$ terms.
Furthermore, the span of our sea quark masses are unable to provide
any meaningful constraint on terms involving the sea masses.
In particular, we neglect $c_2$ as ${\cal O}(\delta m^2)$ and $c_4$ as
${\cal O}(\alpha\delta m)$.
The $c^{EM}$ terms could be determined with simulations at different
values of the QED gauge coupling, however in our present study these
terms are simply absorbed into a redefinition of the relevant
expansion parameters to give
 \begin{align}
M^2(a \bar b) =& M_0^2 + \alpha ( \delta \mu_a + \delta \mu_b) + \beta_1 ( \delta \mu_a^2 + \delta \mu_b^2)   \nonumber
\\&+ \beta_2 ( \delta \mu_a - \delta \mu_b)^2  + \beta_1^{EM} (e_a^2 + e_b^2) + \beta_2^{EM} (e_a- e_b)^2  \nonumber
\\&+\gamma_1^{EM} ( e_a^2 \delta \mu_a + e_b^2 \delta \mu_b ) +\gamma_2^{EM} (e_a e_b) (\delta \mu_a + \delta \mu_b ) \nonumber 
\\&+\gamma_3^{EM} ( e_b^2\delta \mu_a + e_a^2\delta \mu_b )\,.
\label{eq:meson_expansion} 
\end{align}

To the same order in the flavour-breaking parameters, we write the
general expressions for the octet baryons:
\begin{align}
M(aab)=& M_0  +\alpha_1(2\delta \mu_a+\delta \mu_b)+\alpha_2\delta \mu_a \nonumber
\\&+\beta_1 ( 2\delta \mu_a^2 + \delta \mu_b^2)+\beta_2(\delta \mu_a^2+2\delta \mu_a\delta \mu_b) +\beta_3(\delta \mu_a^2)\nonumber
\\&+\beta_1^{EM} (2e_a^2 + e_b^2)+\beta_2^{EM} (e_a^2+2e_ae_b)  +\beta_3^{EM}(e_a^2) \nonumber
\\&+\gamma_1^{EM} ( 2e_a^2 \delta \mu_a + e_b^2 \delta \mu_b )
+\gamma_2^{EM} \left[2\delta \mu_a e_a(e_a +e_b)+2\delta \mu_b e_be_a))\right] \nonumber
\\&+\gamma_3^{EM} (2\delta \mu_a e_a e_b + \delta \mu_b e_a^2) \nonumber
+\gamma_4^{EM} (2\delta \mu_a (e_a^2+e_b^2)+ 2\delta \mu_b e_a^2 ) \nonumber
\\&+\gamma_5^{EM} \delta \mu_ae_a^2+ \gamma_6^{EM} \delta \mu_a
e_ae_b, \label{eq:octet_expansion}
\end{align}
and the decuplet baryons:
\begin{align}
M(abc)=&M_0 +\alpha_1(\delta \mu_a+\delta \mu_b+\delta \mu_c) \nonumber
\\&+\beta_1 ( \delta \mu_a^2 + \delta \mu_b^2+\delta \mu_c^2 ) +\beta_2(\delta \mu_a\delta \mu_b+\delta \mu_a\delta \mu_c +\delta \mu_b \delta \mu_c) \nonumber
\\&+\beta_1^{EM} (e_a^2 + e_b^2+e_c^2)  + \beta_2^{EM} ( e_a e_b + e_a e_c + e_b e_c) \nonumber
\\&+\gamma_1^{EM} ( e_a^2 \delta \mu_a + e_b^2 \delta \mu_b + e_c^2 \delta \mu_c) \nonumber
\\&+\gamma_2^{EM} \left[ \delta \mu_a e_a (e_b+e_c)+ \delta \mu_b e_b(e_a+e_c)+ \delta \mu_c e_c (e_a+e_b) \right] \nonumber
\\&+\gamma_3^{EM} (\delta \mu_a e_b e_c +\delta \mu_b e_a e_c + \delta \mu_c e_a e_b) \nonumber
\\&+\gamma_4^{EM} \left[ \delta \mu_a (e_b^2 +e_c^2) + \delta \mu_b (e_a^2 +e_c^2)+ \delta \mu_c (e_a^2+e_b^2) \right] . \label{eq:decuplet_expansion}
\end{align}
As argued above, we have already dropped the terms involving the sea
quark masses and charges.
We note that for any $f(M)$ an SU(3) flavour and charge breaking
expansion can be made.
For heavy quark masses, due to curvature in the numerical data,
it was found \cite{Horsley:2014koa} to be advantageous to expand
$M^2$; here as the quark mass range used is smaller it is sufficient
to consider an expansion of $M$.

%
%%%%%
%%%%%%%%%%%%%%%%%%%%%%%%%%%%%%%%%%%%%%%%%%%%%%%%%%%%%%%
%%%%%
%

\section{Lattice matters}
\label{sec:lattice matters}
The QCD+QED action we are using in this study is given by
\begin{align}
  S=S_G+S_A+S^u_F+S^d_F+S^s_F
  \label{eq:action}
\end{align}
where $S_G$ is the tree-level Symanzik improved SU(3) gauge action;
$S_A$ is the noncompact U(1) gauge action of the photon; and $S_F^q$
is the fermion action for each quark flavour, $q$.
The photon action is, 
\begin{align}
S_A = \frac{1}{2e^2} \sum_{x,\mu<\nu}
(A_\mu(x)+A_\nu(x+\mu)-A_\mu(x+\nu)-A_\nu(x))^2\,.
\end{align}
For the fermion action we employ the nonperturbatively
$\mathcal{O}(a)$-improved SLiNC action \cite{Cundy:2009yy}
\begin{align}
S_F^q =& \sum_x \left\{ \frac{1}{2} \sum_\mu \left[
  \bar{q}(x)(\gamma_\mu-1)e^{-ie_q A_\mu(x)} \tilde{U}_\mu(x)q(x+\hat
      {\mu}) \right. \right. \nonumber \\
      & \left. \left.- \bar{q}(x)(\gamma_\mu+1) e^{ie_qA_\mu(x)}
      \tilde{U}_\mu^\dagger(x- \hat{\mu})q(x- \hat{\mu})\right]
\right. \nonumber \\
&  \left. + \frac{1}{2 \kappa_q} \bar{q}(x)q(x)-\frac{1}{4}c_{SW}
\sum_{\mu \nu} \bar{q}(x)\sigma_{\mu \nu} F_{\mu \nu} q(x) \right\}
\end{align}
where $\tilde{U}_\mu$ is a single-iterated mild stout-smeared
link.
The clover coefficient $c_{SW}$ has been computed non-perturbatively
for pure QCD \cite{Cundy:2009yy} and we do not include the QED clover
term.

Simulations are carried out on lattice volumes of size $32^3\times 64$
and $48^3 \times 96$.
The sea quark $\kappa$ values are shown in \Tref{tab:lattice}, using charges
of $e_u=+2/3$, $e_d=e_s=-1/3$.
\begin{table}[t]
%\centering
  %\begin{ruledtabular}
  \caption{Summary  of lattice ensemble details. \label{tab:lattice}}
  \begin{indented}
    \lineup
\item[]\begin{tabular}{ccccccc}
\br
$\beta$& $e^2$ & V & $\kappa_u, ~+2/3$ & $\kappa_d, ~-1/3$ & $\kappa_s, ~-1/3$ & Ensemble\\
\mr
5.50 & 1.25& $32^3\times 64$ & 0.124362 & 0.121713 & 0.121713  & 1\\
5.50 & 1.25& $32^3\times 64$ & 0.124440 & 0.121676 & 0.121676  & 2\\
5.50 & 1.25& $32^3\times 64$ & 0.124508 & 0.121821 & 0.121466  & 3\\
5.50 & 1.25& $48^3 \times 96$ & 0.124362 & 0.121713 & 0.121713 & 4 \\
5.50 & 1.25& $48^3 \times 96$ & 0.124440 & 0.121676 & 0.121676 & 5 \\
\br
\end{tabular}
\end{indented}
\end{table}
The strong coupling was chosen to be $\beta=5.50$ and the
electromagnetic coupling was chosen to be $e^2=1.25$, about ten times
greater than physical.
These choices lead to a lattice spacing of
$a=0.068(1)$fm~\cite{Horsley:2015vla}.
Further details can be found in Refs.~\cite{Horsley:2015vla,Horsley:2015eaa}.
In order to better constrain the {\em a priori} unknown coefficients
in the flavour-breaking expansions, we employ up to eight different
partially-quenched valence quarks corresponding to neutral
pseudoscalar meson masses in the range $225\,\text{MeV}\lesssim
M(q\bar{q})\lesssim 765\,\text{MeV}$ and valence quark charges
$e_{a,b}=0,-1/3,+2/3$.
Hadron correlators are evaluated in the so-called $\text{QED}_L$
formulation \cite{Hayakawa:2008an}, where the zero mode of the
photon field is eliminated on each time slice before computing the
valence quark propagators.

Hadron masses are computed from two-point correlation functions using
conventional techniques.
In particular, for baryons we construct zero-momentum two-point
functions as
\begin{equation}
  C(t) = \sum_{\vec{x}}\text{Tr}\,\Gamma
  \left\langle \chi(\vec{x},t)\bar{\chi}(0)\right\rangle,
  \label{eq:2pt}
\end{equation}
for some choice of baryon spin projection matrix, $\Gamma$,
e.g. for spin averaged, $\Gamma = (1+\gamma_4)/2$.
For octet baryons, we employ the interpolating operator in terms of a
doubly-represented quark of flavour, $q_1$, and a singly-represented
quark of flavour, $q_2$
\begin{equation}
\chi(\vec{x},t) = \epsilon^{abc}\big(
q_1^{aT}(\vec{x},t)\,C\gamma_5\,q_2^b(\vec{x},t)\big)
q_1^c(\vec{x},t)\ ,
\end{equation}
where here $a,b,c$ are colour labels.
In the following, given the partially quenched nature of our
simulations, we distinguish flavour by the electric charge carried by
a quark rather than its mass.
For example, when the combination $uud$ occurs in the following
discussion, this refers to an octet baryon where its
doubly-represented quark has charge $+2/3$ while the
singly-represented quark has charge $-1/3$.
For decuplet baryons we choose an explicit spin-projection for the
scalar diquark of the interpolating operator that contains doubly- and
singly-represented quarks
\begin{align}
  \chi(\vec{x},t) = \frac{1}{\sqrt{3}}\epsilon^{abc}\Big[
    &2 \big( q_1^{aT}(\vec{x},t)\,C\gamma_-\,q_2^b(\vec{x},t) \big)
    q_1^c(\vec{x},t) \nonumber\\
    &+ \big( q_1^{aT}(\vec{x},t)\,C\gamma_-\,q_1^b(\vec{x},t) \big)
q_2^c(\vec{x},t)\Big]\ ,
\end{align}
where $\gamma_- = (\gamma_2 + i\gamma_1)/2$.
We note that correlation functions for the $\Sigma^{*0}$, involving
all three flavours of quarks, have not been computed in the present study.

In Figs.~\ref{fig:octetspectrum} and \ref{fig:decupletspectrum} we
show examples of our results for the octet and decuplet
baryon masses, respectively.
\begin{figure*}[h]
\begin{indented}
\item[]\includegraphics[width=0.8\textwidth]{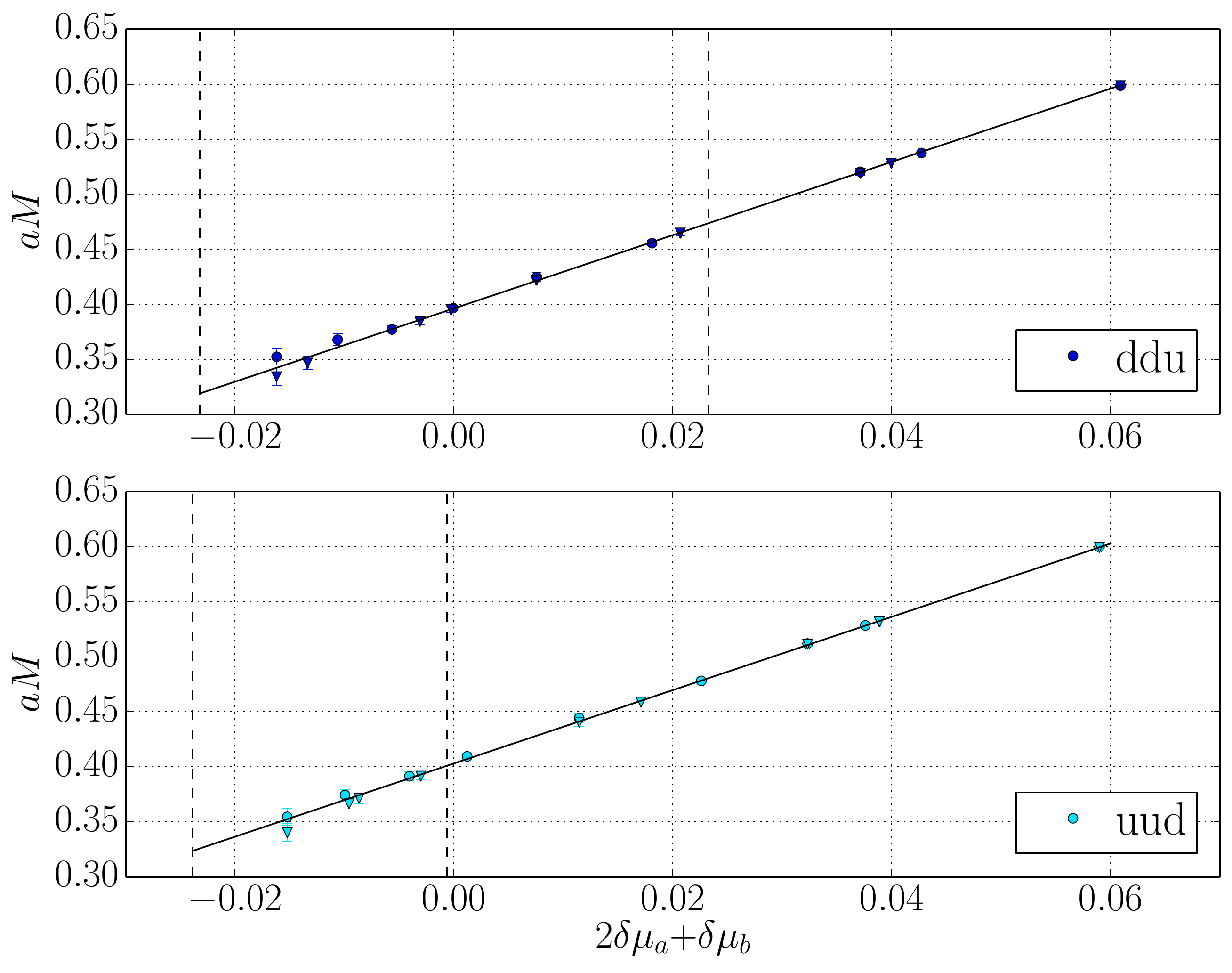}
  \caption{The dependence of the neutral, $Q=0$, (top) and positively-charged,
    $Q=+1$, (bottom) octet baryons on the leading SU(3) breaking term
    $(2\delta\mu_a + \mu_b)$ from Eq.~(\ref{eq:octet_expansion}), as
    described in the text.
    The circles and triangles denote results from the $48^3\times 96$
    ensembles 4 and 5, respectively, as labelled in
    \Tref{tab:lattice}.
    The vertical dashed lines indicate the locations of the baryon
    masses at the physical quark masses, as given in
    Table~\ref{tab:quarkmasses}.
    \label{fig:octetspectrum}}
\end{indented}
\end{figure*}
These figures collect the partially-quenched results on both
$48^3\times 64$ ensembles listed in Table~\ref{tab:lattice}, plotted
as function of the leading SU(3) breaking term $(2\delta\mu_a +
\mu_b)$ from Eq.~(\ref{eq:octet_expansion}).
The circles and triangles denote results from the $48^3\times 96$
ensembles 4 and 5, respectively, as labelled in \Tref{tab:lattice}.
To facilitate simpler comparisons with the fitted SU(3) expansion, the
contributions from all other quark mass-dependent terms beyond
$(2\delta\mu_a + \mu_b)$ in Eq.~(\ref{eq:octet_expansion}) have been
subtracted from each mass point.
As such the curves displayed are given by:
\begin{equation}
M^{(sub)}(aab)=M_0 +\alpha_1(2\delta \mu_a+\delta \mu_b)
+\sum \beta_i^{EM}F_i(e_a,e_b),
\end{equation}
where the functions $F_i$ encode the appropriate quark-charge
dependence as given by Eqs.~\ref{eq:octet_expansion} and
\ref{eq:decuplet_expansion}.
The residual scatter of the points around the line provides an indication of the quality of
the global fit across the baryons.
The slight difference in between the lines of the different panels
provides a measure of the QED splittings encoded by the $\beta^{EM}$ terms.

The left vertical dashed lines in Figs.~\ref{fig:octetspectrum} and
\ref{fig:decupletspectrum} display the positions of the physical
point for the corresponding baryons.
For the octet baryons at charge $Q=0$, the fit lines ``interpolate'' to
the $\Xi^+$ and a mild extrapolation to the neutron point.
Similarly, we see the $\Sigma^+$ and proton in the charge $Q=1$ panel.
For the decuplet we have chosen to display the corresponding $Q=-1$
and $Q=+1$ contours.

\begin{figure*}[ht!]
\begin{indented}
\item[]\includegraphics[width=0.8\textwidth]{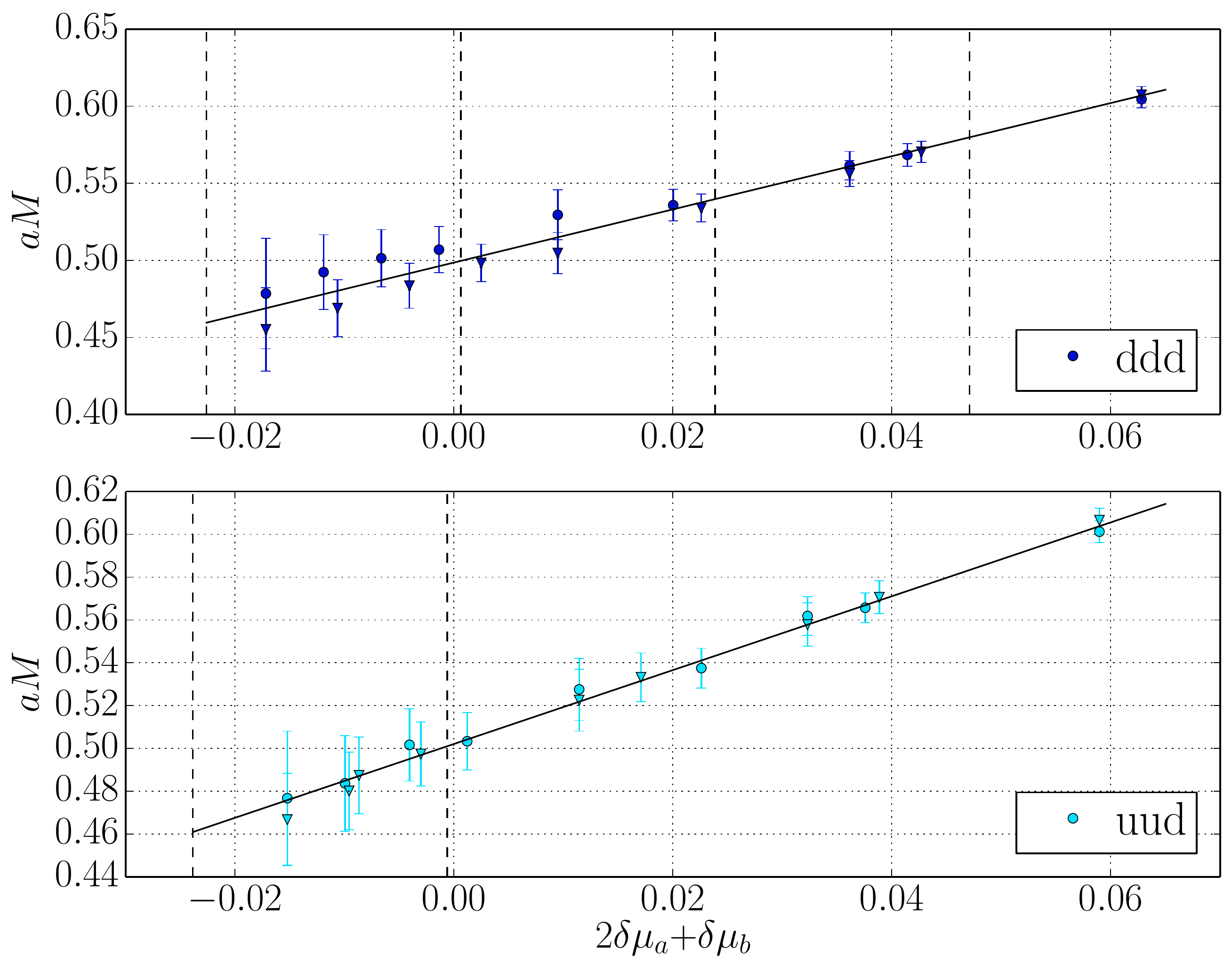}
\end{indented}
\caption{The dependence of the negative-charge, $Q=-1$, (top) and positively-charged,
    $Q=+1$, (bottom) decuplet baryons on the leading SU(3) breaking term
    $(2\delta\mu_a + \mu_b)$ from Eq.~(\ref{eq:decuplet_expansion}),
  as described in the text.
    The circles and triangles denote results from the $48^3\times 96$
    ensembles 4 and 5, respectively, as labelled in
    \Tref{tab:lattice}.
    The vertical dashed lines indicate the locations of the baryon
    masses at the physical quark masses, as given in
    Table~\ref{tab:quarkmasses}.
\label{fig:decupletspectrum}}
\end{figure*}

%%%%%%%%%%%%%%%%%%%%%%%%%%%%%%%%%%%%%%%%%%%%%%%%%%%%%%%
%%%%%%%%%%%%%%%%%%%%%%%%%%%%%%%%%%%%%%%%%%%%%%%%%%%%%%%
\section{Analysis}
\label{sec:determining}

\subsection{Finite volume}
In the present work, we just consider the leading finite-volume
corrections associated with the electromagnetic interactions.
Strong interaction effects are expected to be subdominant as they are
exponentially suppressed by $\exp(-m_\pi L)$ whereas electromagnetic
finite volume (FV) corrections are only power law suppressed,
i.e. $1/L^n$.
These infinite volume hadron masses are estimated in the effective
field theory, NRQED$_L$, including corrections up to (and including)
${\cal O}(1/L^2)$
\cite{Borsanyi:2014jba,Davoudi:2014qua,Lee:2015rua}.\footnote{We note
  that with our larger value of the electromagnetic coupling, $\alpha_{\rm
    QED}\sim 0.1$, this expansion parameter is numerically comparable
  to the values of $1/(mL)$ in the NRQED expansion
  \cite{Matzelle:2017qsw}.
Nevertheless, Matzelle and Tiburzi \cite{Matzelle:2017qsw} have shown
that potentially-relevant higher-order terms in $\alpha_{\rm QED}$ do
not affect the expansion to ${\cal O}(1/L^2)$.}

Our full dataset includes results obtained from a subset of
simulations which have all simulation parameters fixed except the
physical volume.
This allows us to assess the effects of the finite volume
on our simluations as compared to the analytic expectations of
Refs.~\cite{Borsanyi:2014jba,Davoudi:2014qua,Lee:2015rua}.
By considering mass differences between isospin partners,
strong finite-volume effects should cancel, leaving us with quantites
that are primarily sensitive to electromagnetic finite-volume effects.
We find that the splittings on our two volumes ($\sim 2.2, 3.3$~fm) are generally compatible with each other after accounting for the leading QED FV effects. 
When quoting our final values in the following sections, the results
obtained from the finite-volume corrected $48^3\times 96$
lattice data provide the central values and statistical uncertainties.
The difference between the two volumes, after
correcting for the leading-order EM finite-volume effects and extrapolating to the physical point, provides a
conservative estimate for the dominant systematic uncertainty.

\subsection{The physical point}

The first stage of our analysis is to identify the location of the
quark mass parameters corresponding to the physical point.
For this, we restrict ourselves to the meson sector following the
procedure outlined in Ref.~\cite{Horsley:2015vla}.
The general expansion of Eq.~(\ref{eq:meson_expansion}) is modified
such that the QED contributions to the neutral pseudoscalar mesons are
absorbed into the quark self-energy.
This modification defines the Dashen quark mass parameters, $\delta
m^D_q,\, \delta\mu_a^D$, which are then used to parameterise the
deviation from the SU(3) origin.
As described in \Sref{sec:massexpansion}, terms involving the
$c$ and $c^{EM}$ coefficients have been neglected in this analysis.
Upon fitting the resulting expression to the remainder of the
pseudoscalar meson mass spectrum, the enhanced value of
  $\alpha_{\text{QED}}=1.25/4\pi$ employed in our simulations is corrected by a
  linear rescaling of the fitted $\beta^{EM}$ and $\gamma^{EM}$
  coefficients by a factor of $4\pi/(1.25\times 137)$.
Constraining the fits to three pieces of physical input, namely the
physical $\pi^0,\ K^0$ and $K^+$ masses, then leads to a determination
of the lattice spacing and the bare quark masses at the physical
point.
These results are given in \Tref{tab:quarkmasses} for the larger
$48^3\times 96$ volume.
We note that only three physical inputs are required to
determine the four unknown parameters, as we have the additional
constraint built into our simulations that the average quark mass,
$\bar{m}=(m_u+m_d+m_s)/3$, is held fixed, i.e. $\delta m_u + \delta
m_d + \delta m_s = 0$.
Using the parameters given in \Tref{tab:quarkmasses}, we are able to
provide a prediction for the $\pi^+$ mass, which is provided in
\Tref{tab:octetsplit} in the form of a mass splitting from the
$\pi^0$.
The result from the present work is in agreement with that from
\cite{Horsley:2015vla}, however we note the improved statistical
precision of the current work due to the inclusion of the additional
ensembles away from the SU(3) symmetric point summarised in
\Tref{tab:lattice}.

\begin{table}[h]
\caption{Dashen quark mass parameters at the physical point and the
  inverse lattice spacing. \label{tab:quarkmasses}}
\begin{indented}
\lineup
\item[]\begin{tabular}{cccc}
\br
$a\delta m^D_u$ & $a\delta m^D_d$ & $a\delta m^D_s$ & $a^{-1}$/GeV\\
 -0.00786 (1) & -0.00728 (2) & 0.0151 (2) & 2.906 (12) \\
\br
\end{tabular}
\end{indented}
\end{table}

\begin{table}[t!]
\caption{Predicted mass splittings for $\pi^+$ and octet baryons in
  the Dashen scheme, including a separation into QCD and QED
  contributions in the Dashen scheme. $\pi^0$ assumed to be the state
  $(u \bar u - d \bar d)/\sqrt 2$. Experimental mass splittings
  \cite{Patrignani:2016xqp} are also given for
  comparison.
  All values quoted in MeV.\label{tab:octetsplit}}
\begin{indented}
\lineup
\item[]\begin{tabular}{ccccc}
\br
&$\pi^+-\pi^0$     & $n-p$          & $\Sigma^- - \Sigma^+$ & $\Xi^- - \Xi^0$ \\
\mr
QED   & 5.86(14)(40) & $-1.53(25)(50)$ & $-0.29(24)(10)$ & 1.19(15)(20) \\
QCD   & ---          &   2.79(67)(40)  &   8.58(72)(70)  & 5.79(28)(80) \\
Total & ---          &   1.27(75)(50)  &   8.29(77)(25)  & 6.95(25)(90) \\
\mr
Experiment & 4.59 & $1.30$ & 8.08 & 6.85\\
\br
\end{tabular}
\end{indented}
\end{table}

\subsection{Baryons}

At this stage we have completely described our Dashen scheme and have
predictions for the physical quark masses and the lattice spacing
for each volume.
Hence we are now in a position to fit the finite-volume corrected,
partially quenched octet and decuplet baryon masses to the
flavour-breaking expansions given in Eqs.~(\ref{eq:octet_expansion})
and (\ref{eq:decuplet_expansion}) with the bare quark masses, $\delta
\mu_q$ replaced by the Dashen mass $\delta \mu_q^D$.

Previous work has shown that the light hadron spectrum in pure QCD is
well described along our $\bar{m}=\text{constant}$ trajectory by
flavour breaking expansions that are linear in the flavour breaking
quark mass parameter over the entire mass range from the
SU(3)-symmetric point to the physical point, with only small
corrections provided by terms quadratic in the flavour-breaking
parameter \cite{Bietenholz:2011qq}.
A summary of the fit parameters for the $48^3\times 96$ lattice
ensembles is presented in \ref{sec:fitpars}.

We note that the reduced $\chi^2$ values indicate that the fits are
suitably able to describe the data. 
To visualise the multi-dimensional fit, we show the
central values of the fit parameterisation against the 
(finite-volume corrected) lattice spectra
in Figs.~\ref{fig:octetspectrum} and
\ref{fig:decupletspectrum}.
%

%
%%%%%
%%%%%%%%%%%%%%%%%%%%%%%%%%%%%%%%%%%%%%%%%%%%%%%%%%%%%%%
%%%%%
%

%%%%%%%%%%%%%%%%%%%%%%%%%%%%%%%%%%%%%%%%%%%%%%%%%%%%%%%
%%%%%%%%%%%%%%%%%%%%%%%%%%%%%%%%%%%%%%%%%%%%%%%%%%%%%%%
\section{Results \& Discussion}
\label{sec:results}

%%%%%%%%%%%%%%%%%%%%%%%%%%%%%%%%%%%%%%%%%%%%%%%%%%%%%%%
\subsection{Octet baryons}
Using the preferred fits we can extrapolate our spectrum to the
physical point, as determined within the meson sector.
The absolute masses of the baryon octet are summarised in
\Tref{tab:octetmass}, where we see excellent agreement with the
experimental values for the proton and neutron masses, while we
observe multiple-$\sigma$ discrepancies as the number of strange
quarks in the baryons is increased.
This is perhaps an indication of a slight mismatch in our tuning of
the singlet quark mass.
This effect, however, will not affect the results for isospin
splittings presented in the remainder of this paper.
\begin{table}[t!]
%  \begin{flushleft}
\caption{Extrapolated masses for the octet baryons in the Dashen scheme,
  showing comparison with the experimental masses
  \cite{Patrignani:2016xqp}. Only the
  maximally-charged state of each isospin multiplet is shown.
  All values quoted in MeV. \label{tab:octetmass}}
\begin{indented}
\lineup
\footnotesize
\item[]\begin{tabular}{cccc}
\br
           &$p$          & $\Sigma^+$   &$\Xi^0$    \\
\mr
           & 939(14)(56) & 1165(11)(23) & 1276(6)(19)  \\
\mr
Experiment & 938.3       & 1189.4       & 1314.8   \\
\br
\end{tabular}
%\end{flushleft}
\end{indented}
\end{table}

Given the high degree of correlation in the mass determinations, the
isospin splittings are determined to much better precision and are
displayed in \Fref{fig:octetRelsplittingD} for our two lattice volumes.
\begin{figure}[tb!]
\begin{indented}
\item[]\includegraphics[width=.7\textwidth]{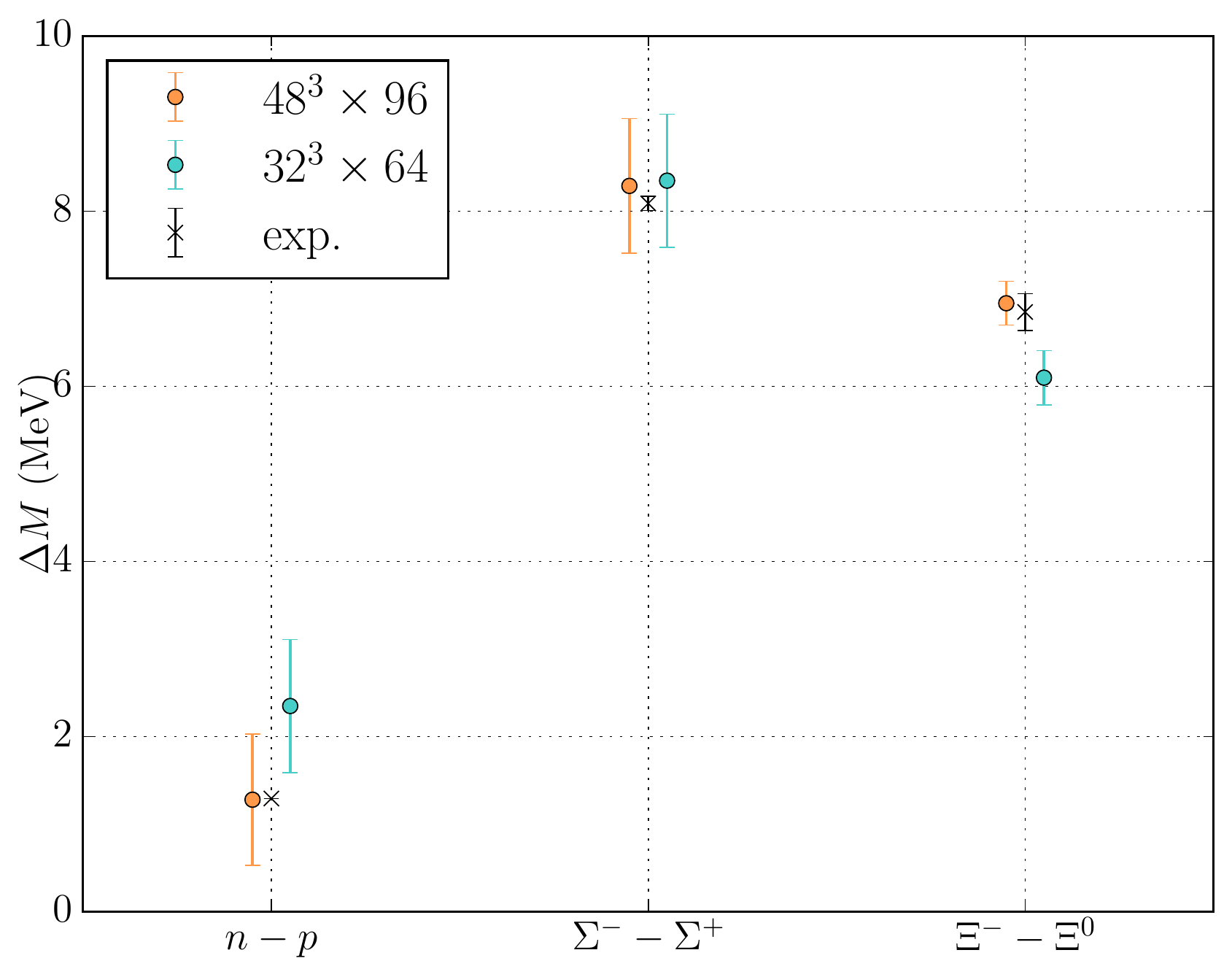}
\end{indented}
\caption{Octet mass splitting with the average octet family mass
  subtracted. This includes EM effects. The black crosses are
  experimental data. The coloured points are estimates generated from
  our lattice analysis.
\label{fig:octetRelsplittingD}}
\end{figure}

These mass splittings are summarised in \Tref{tab:octetsplit} and
serve as an update to our earlier work \cite{Horsley:2015eaa} which
was based on only a single set of sea quark masses, namely ensembles 1
and 4 in \Tref{tab:lattice}.
We note that in Ref.~\cite{Horsley:2015eaa} the photon zero modes were treated dynamically, requiring an effective kinetic energy to be subtracted at the analysis stage.
The first uncertainty shown in \Tref{tab:lattice} is statistical, while the
second provides an estimate of the finite-size systematic error as
described in the previous section.
We note that since our simulations are performed at only a single
value of the lattice spacing, no continuum extrapolation is possible. 
As a guide to the magnitude of these UV cut-off effects, Ref. \cite{Durr:2008ail,Durr:2009ssd} showed (with lattice spacings of a similar size and similar quark and gluon actions) that corrections to the nucleon and $\Delta$ mass are of the order $1\,$\%. Additionally, Ref. \cite{Borsanyi:2014jba} used a similar lattice spacing and action as in the present work and showed that the UV cut-off effects on the isospin mass splittings of the octet baryons, including QED, were on the order of $1\%$.

In Ref.~\cite{Horsley:2015vla} we provided a prescription for
converting electromagnetic mass contributions between Dashen and
$\overline{\text{MS}}$ schemes, however to leading order this has no
effect on the central values and hence we only quote our Dashen scheme
results.
To separate QED and QCD, we note the $\gamma^{EM}$ terms in Eqs.~\ref{eq:octet_expansion} and
  \ref{eq:decuplet_expansion} describe a product of $e^2$ and
  $\delta\mu$ effects.
  We distinguish the
  isospin-breaking effects arising from these terms
  as either being: QED, when $\delta\mu_u=\delta\mu_d$; QCD, when
  $e_u=e_d$; or a remaining (and small) second-order isospin-breaking
  effect. For example, terms involving the product
  $(e_u-e_d)(\delta\mu_u+\delta_d)$ is attributed to QED, whereas
  $(e_u+e_d)(\delta\mu_u-\delta_d)$ is attributed to QCD. The former
  vanish if the up and down charges are equal, while the latter
  vanish if the masses are equal.

The electromagnetic splitting between the proton and neutron has seen
considerable attention in recent years.
Our result for the proton-neutron mass-splitting shows some preference to the
dispersive analysis of Ref.~\cite{WalkerLoud:2012bg}, which finds
$-1.30\pm 0.47\mev$.
In contrast, we see our result is slightly larger in magnitude than
the values reported in 
Refs.~\cite{Erben:2014hza} and \cite{Gasser:2015dwa,Gasser:1974wd},
though not in statistical disagreement.
It is noted that the latter phenomenological studies display better agreement with the
lattice results of the BMW Collaboration \cite{Borsanyi:2014jba}.
\Fref{fig:OctetemisosplittingD} shows the breakdown
between strong isospin breaking and electromagnetic
effects for the octet baryons.
The $\Xi$ splittings are generally
compatible with both phenomenological estimates \cite{Erben:2014hza}
and the BMW lattice results \cite{Borsanyi:2014jba}.
We note that a direct comparison for the electromagnetic splitting in the $\Sigma$ is not possible, since this is
set to zero in the scheme prescribed in Ref.~\cite{Borsanyi:2014jba}.
\begin{figure}[ht!]
\begin{indented}
\item[]\includegraphics[width=.8\textwidth]{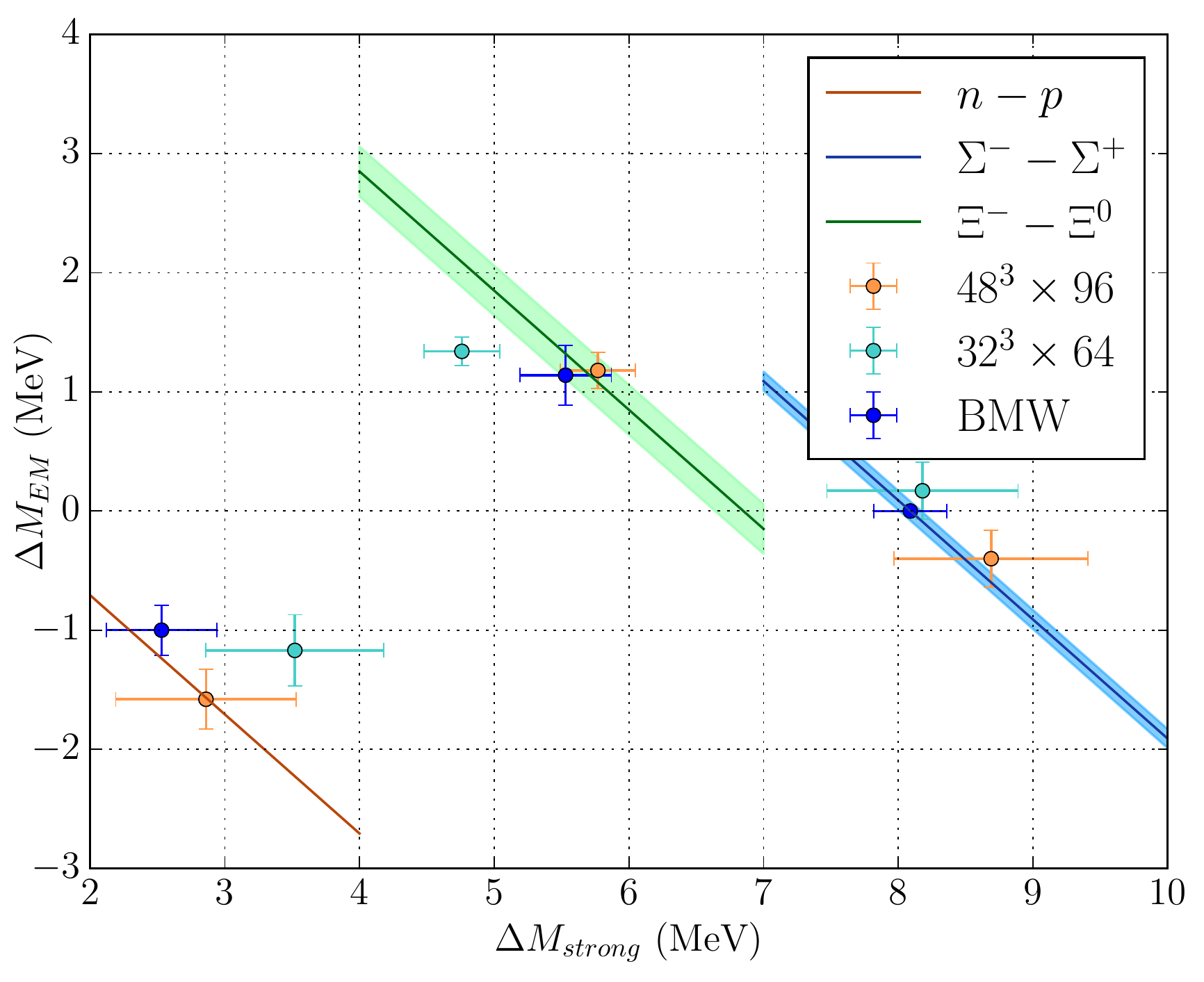}
\end{indented}
\caption{A decomposition of the octet splitting in terms of EM and
  strong isospin breaking effects. The BMW lattice points are from
  \cite{Borsanyi:2014jba}. The lines represent a constraint placed by
  the experimental data.
\label{fig:OctetemisosplittingD}}
\end{figure}

%%%%%%%%%%%%%%%%%%%%%%%%%%%%%%%%%%%%%%%%%%%%%%%%%%%%%%%
\subsection{Decuplet baryons}

For the decuplet baryons, our analysis is restricted to the
extrapolation of our lattice masses to the physical point based on the
flavour-breaking expansion about the SU(3) symmetric point.
That is, no attempt has been made in the present work to incorporate
the effects of the resonant nature of the decuplet baryons at the
physical quark masses --- which necessarily lead to branch point
singularities in the quark mass extrapolation
\cite{Young:2002cj,Pascalutsa:2005nd}.
The isospin splittings within the decuplet baryons therefore represent
a first estimate on the magnitude of these effects.
A full treatment including mixing with multi-hadron states is left for
future work. 

As for the octet baryons, our decuplet expansion is fit to the
electromagnetic finite-volume corrected lattice results.
The absolute masses themselves do not compare so favourably with
experimental determinations, as shown in \Tref{tab:decupletmasses}.
\begin{table}[h]
\caption{Absolute masses for the maximally-charged state for each
  isospin multiplet within the decuplet.
  All values quoted in MeV.\label{tab:decupletmasses}}
\begin{indented}
\lineup
\item[]\begin{tabular}{lcccc}
\br
           & $\Delta^{++}$ & $\Sigma^{*+}$ & $\Xi^{*0}$ & $\Omega$\\
\mr
This work  & 1304(59)(6)      & 1425(38)(8)      & 1542(26)(9)   & 1656(21)(8)  \\
\mr
Experiment & $1231$        & $1383$        & $1532$     & $1672$\\
\br
\end{tabular}
\end{indented}
\end{table}
Within the quoted uncertainties we observe that the
  absolute masses at the physical point are compatible with
  the experimental
  masses.
Nevertheless, it is possible that there is a systematic uncertainty
that is causing an underestimate of the overall scale of the SU(3)
breaking between these states.
This could be due to the fact that our simulations are performed at
and around the SU(3) symmetric point where the $\Delta$ and
$\Sigma^*$ states are stable states.
However, in the physical system the $\Delta$ and $\Sigma^*$ states are
unstable and decay, e.g. to $\Delta \to \pi+N$, where the net mass of
the $\pi+N$ system is significantly lower than the three quark state.
The opening of these decay channels is certainly anticipated to affect
the extrapolation to the physical point \cite{Young:2002cj}.
This physics of the decays would become less prominent for $\Xi^*$ and irrelevant for
the $\Omega$, which is stable under the strong interaction. 
Based on analysis in the literature \cite{Young:2002cj,Pascalutsa:2005nd,Aceti:2014xpy} we expect more favourable agreement when these effects are taken into account. For instance,  using the physical decuplet masses as input and chiral perturbation theory Ref. \cite{Aceti:2014xpy} and Ref. \cite{Pascalutsa:2005nd} give estimates of the masses of the decuplet baryons when these decay channels are turned off.  These estimates match more closely with our lattice mass predictions.

Assuming that the threshold effects do not have a strong influence on
the isospin-violating parts, we expect that the magnitudes and
orderings of the splittings to be indicative of the expected behaviour
at the physical point.
We highlight some various selected splittings of phenomenological
interest in \Tref{tab:decupletsplit}.
The combination $\Delta^{++}+\Delta^{-}-\Delta^{+}-\Delta^{0}$ is
selected as it eliminates the leading strong isospin violation, and
hence provides a purely electromagnetic effect.
The difference $\Delta^0-\Delta^{++}$ is reported by the PDG.
The particular combination
$\Delta^--\Delta^{++}+\tfrac13\left(\Delta^0-\Delta^+\right)$ can be
isolated experimentally by considering the difference between $\pi^+$
and $\pi^-$ cross sections on deuteron targets, as reported in
Ref.~\cite{Pedroni:1978it}.
For the $\Sigma^*$ baryons, again
$\Sigma^{*+}+\Sigma^{*-}-2\Sigma^{*0}$ removes the leading strong
isospin breaking, leaving a purely electromagnetic effect.
To mimic the analogous splitting of the octet baryons we display
$\Sigma^{*-}-\Sigma^{*+}$, which is observed to be dominated by the
quark mass differences.
Similarly we find that $\Xi^{*-}-\Xi^{*0}$ is also dominated by
the strong effect, which is perhaps counterintuitive since the
electromagnetic effect in $\Xi^{*-}$ is repulsive while it is
attractive in $\Xi^{*0}$.

\begin{table}[t!]
\lineup
\caption{Mass splittings for decuplet baryons.
All values quoted in MeV. \label{tab:decupletsplit}}
\footnotesize
%\item[]
\begin{tabular}{lccc}
\br
& $\Delta^{++}+\Delta^{-}-\Delta^+-\Delta^0$ & $\Delta^0-\Delta^{++}$
& $\Delta^--\Delta^{++}+\tfrac13 \left(\Delta^0-\Delta^+\right)$\\
\mr
QED & 1.7(14)(10) & -2.5(20)(13) & -2.7(26)(20) \\
QCD & -0.006(11)(6) & 6.3(24)(5) & 10.5(40)(10) \\
Total & 1.7 (14)(10) & 3.8(31)(5) & 7.8(46)(5)  \\
\mr
Cutkosky \cite{Cutkosky:1992nx} & 2.84--3.55    & 0.81--1.53                     & 4.31--4.92\\
Exp./Pheno.                     & ---           & 2.86(30) \cite{Gridnev:2004mk} & 4.6(2) \cite{Pedroni:1978it} \\
\mr
\mr
& $\Sigma^{*+}+\Sigma^{*-}-2\Sigma^{*0}$ & $\Sigma^{*-}-\Sigma^{*+}$&$\Xi^{*-}-\Xi^{*0}$ \\
\mr
QED & 1.5(7)(1) & -0.8(11)(7) & 0.61(51)(60)  \\
QCD  & -0.0032(56)(30) & 6.1(22)(2) & 2.92(98)(1) \\
Total & 1.5(7)(1) & 5.3(23)(10) & 3.54(98)(8)  \\
\mr
Cutkosky \cite{Cutkosky:1992nx} & 1.42             & 4.56       & 3.09 \\
PDG \cite{Patrignani:2016xqp}   & 2.6(21)          & 4.4(6)     & 3.2(6)\\
\br
\end{tabular}
\end{table}

\begin{figure}[t!]
\begin{indented}
\item[]
\includegraphics[width=.8\textwidth]{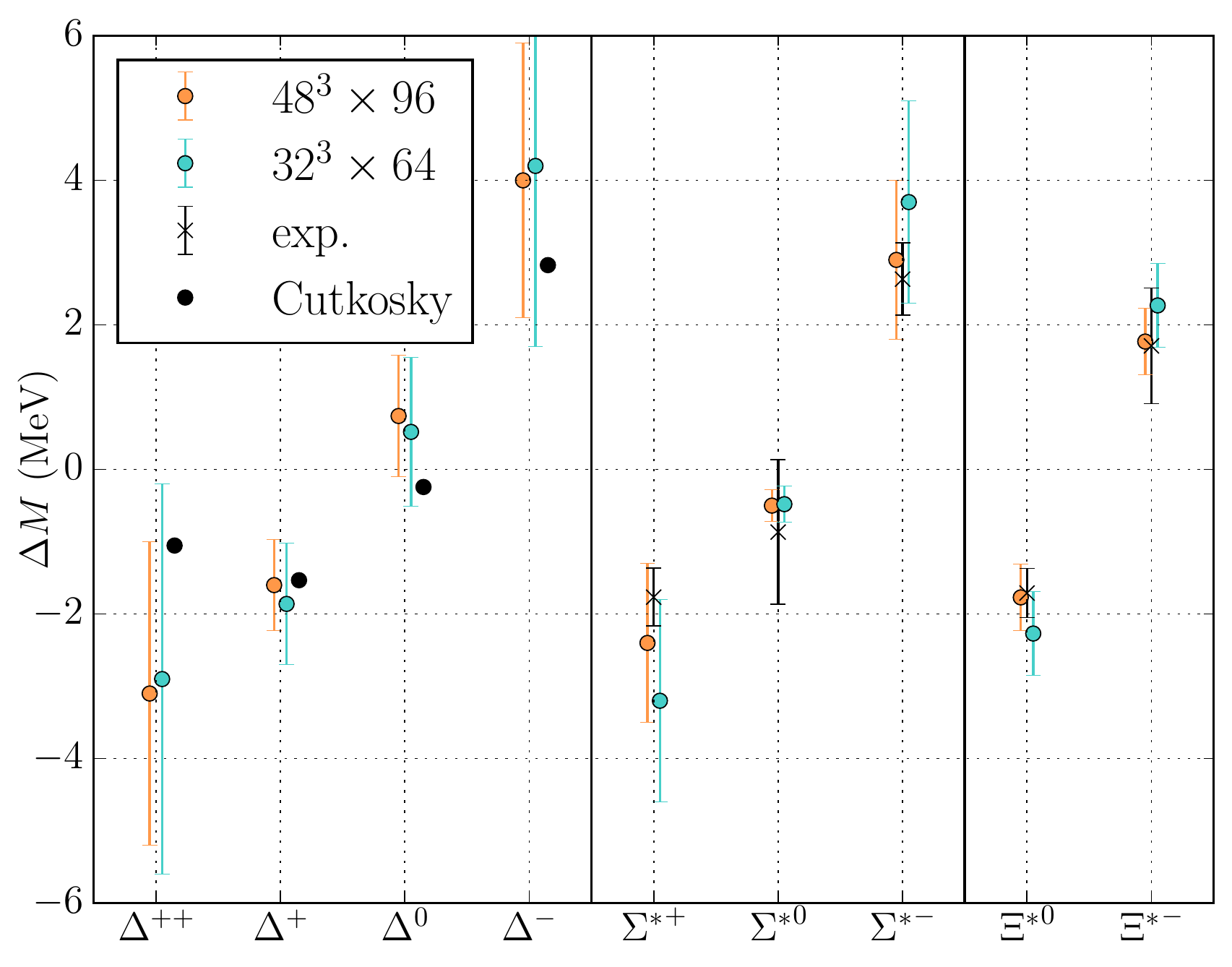}
\end{indented}
\caption{Mass splittings within the isospin multiplets of decuplet
  baryons with respect to the average multiplet mass (see Eq.~(\ref{eq:deltaM})), including both
  strong and electromagnetic effects. The black crosses for $\Sigma^*$
  and $\Xi^*$ baryons are experimental data \cite{Patrignani:2016xqp},
  while the black circles for $\Delta$ baryons indicate a fit to
  experimental data \cite{Cutkosky:1992nx}.
\label{fig:RelsplittingD}}
\end{figure}

The final mass splittings due to isospin breaking effects, both strong
and electromagnetic, at the physical quark masses for all decuplet
baryon on both volumes are shown in \Fref{fig:RelsplittingD}.
The mass splittings within each isospin multiplet are displayed as the
difference of each mass from the average of its respective multiplet.
For example, the splittings in Delta baryons are given by:
\begin{equation}
  \Delta M_B=M_B-\frac14\left(M_{\Delta^{++}}+M_{\Delta^+}+M_{\Delta^0}+M_{\Delta^-}\right) .
  \label{eq:deltaM}
\end{equation}

For the $\Sigma^*$ and $\Xi^*$ baryons we are able to compare our mass
splittings directly with those obtained from experiment, indicated by
the black crosses, while for the $\Delta$ baryons we are only able to
compare to a fit to experimental data \cite{Cutkosky:1992nx}.
The results shown in \Fref{fig:RelsplittingD} clearly agree with the
experimental determinations, indicating that while the overall
magnitude of our decuplet baryon masses are overestimated, potentially
due to the fact that we haven't considered the full resonance
structure of the strongly unstable baryons, the mass splittings within
each multiplet can be accurately described by our QCD+QED simulation.

Finally, we note that the present analysis allows us to estimate the total contributions to baryon masses arising from electromagnetism. Of particular relevance is that the $\Omega$ baryon is now being commonly used to determine the scale in lattice QCD to sub-percent precision. We find the QED contribution to the $\Omega$ mass to be less than 0.2\% of the total mass, below the precision relevant for current lattice QCD simulations, but perhaps significant for the next generation of calculations.

%
%%%%%
%%%%%%%%%%%%%%%%%%%%%%%%%%%%%%%%%%%%%%%%%%%%%%%%%%%%%%%
%%%%%
%

\section{Conclusion}
\label{sec:conclusion}

We have presented lattice QCD+QED results for the light baryon mass
spectrum including both strong and electromagnetic isospin breaking
effects.
Our simulations are based on partially-quenched simulations with 2
volumes and up to three choices for the sea quark masses at and around
the SU(3) symmetric point.
For the octet baryons, this work represents an update to our earlier
findings \cite{Horsley:2015eaa} which were obtained from only a single
choice of sea quarks.
Another difference to our previous work is the use of the $\text{QED}_L$
formulation \cite{Hayakawa:2008an} for the valence quarks.
We find excellent agreement between our results for the mass
splittings of the isospin partners, $n-p,\,\Sigma^- - \Sigma^+,\,\Xi^-
- \Xi^0$ and those observed experimentally.
Our procedure also allows for the decomposition of these
isospin-dependent mass splittings into strong and electromagnetic
contributions with the Dashen scheme.

Qualitatively the absolute values of the masses of our decuplet
spectrum are too large, although we have not yet considered how the
pole position of a resonance can be affected by the multi-hadron
strong decay modes in a finite volume which may account for some of
this discrepancy.
A description of the resonance nature of the decuplet baryon mass
spectrum has only recently started to be addressed in pure QCD lattice
simulations \cite{Alexandrou:2015hxa,Andersen:2017una}.
A full formalism to resolve resonant features of hadron scattering in
a finite box, including the long-range Coulomb
interactions, is yet to be developed.

The principle focus of the present work is the determination of the
isospin breaking effects in the 
decuplet baryon mass spectrum.
The lattice estimates for the mass splittings within the different
isospin multiplets of the decuplet baryons, however, are in excellent
agreement with the experimentally observed splittings in the case of
the $\Sigma^*$ and $\Xi^*$ baryons, or a phenomenological fit using
experimental data \cite{Cutkosky:1992nx} in the case of the $\Delta$
baryons.
%

%
%%%%%
%%%%%%%%%%%%%%%%%%%%%%%%%%%%%%%%%%%%%%%%%%%%%%%%%%%%%%%
%%%%%
%

\ack
The numerical configuration generation (using the BQCD lattice QCD
program \cite{Haar:2017ubh}) and data analysis (using the Chroma
software library \cite{Edwards:2004sx}) was carried out on the IBM
BlueGene/Q and HP Tesseract using DIRAC 2 resources (EPCC, Edinburgh,
UK), the IBM BlueGene/Q (NIC, J\"ulich, Germany) and the Cray XC40 at
HLRN (The North-German Supercomputer Alliance), the NCI National
Facility in Canberra, Australia (supported by the Australian
Commonwealth Government) and Phoenix (University of Adelaide).
HP was supported by DFG Grant No. PE 2792/2-1.  PELR was supported in
part by the STFC under contract ST/G00062X/1 and RDY and JMZ were
supported by the Australian Research Council Grants FT120100821,
FT100100005, DP140103067 and DP190100297.
We thank all funding agencies.

\appendix

\section{Fit parameters}
\label{sec:fitpars}
In Table \ref{tab:fit_coefficients}, we report the fit
parameters of the flavour-breaking expansions.

\begin{table}[h]
\caption{Expansion parameters as determined for the $48^3\times96$
  volume. The terms involving the electromagnetic couplings have been
  scaled to the physical point by the factor $\alpha_{QED}^{\rm
    phys}/\alpha_{QED}^{\rm lat}$.
  \label{tab:fit_coefficients}}
  \begin{indented}
    \lineup
    \item[]
      \begin{tabular}{lccc}
        \br
                  &   Meson          &   Octet        &   Decuplet    \\
\mr
$M_0$             &   0.020504(66)   &   0.3944(24)   &   0.494(11)   \\
$\alpha_1$        &   1.1703(47)     &   3.32(11)     &   1.73(40)    \\
$\alpha_2$        &   ---            & $-1.71(23)$    &   ---         \\
$\beta_1$         & $-0.17(22)$      & $-20.7(40)$    &   1.0(135)    \\
$\beta_2$         &   1.51(12)       & $-14.2(13)$    & $-2.9(45)$    \\
$\beta_3$         &   ---            &   38.0(11)     &   ---         \\
$\beta_1^{EM}$    &   0.0001975(47)  &   0.00083(17)  &   0.00064(53) \\
$\beta_2^{EM}$    & $-0.0005222(37)$ &   0.001032(55) &   0.00042(18) \\
$\beta_3^{EM}$    &   ---            & $-0.00022(33)$ &   ---         \\
$\gamma_1^{EM}$   &   0.00435(26)    & $-0.0041(44)$  &   0.012(17)   \\
$\gamma_2^{EM}$   & $-0.00899(13)$   &   0.0014(11)   &   0.003(5)    \\
$\gamma_3^{EM}$   &   0.00526(21)    & $-0.0063(54)$  &   0.0092(61)  \\
$\gamma_4^{EM}$   &   ---            &   0.0014(27)   & $-0.00011(70)$\\
$\gamma_5^{EM}$   &   ---            &   0.008(11)    &   ---         \\
$\gamma_6^{EM}$   &   ---            &   0.016(12)    &   ---         \\
\mr
$\chi^2$          &   183.74         &   47.12        &   20.35       \\
$DOF$             &   105            &   112          &   118         \\
$\chi^2/DOF$      &   1.75           &   0.42         &   0.172 \\
\br
\end{tabular}
\end{indented}
\end{table}

\break
\bibliographystyle{ieeetr}
\bibliography{deltasplit} 
\end{document}